Dear Reader,

below you find the LaTeX source of our paper PRA-HEP-
94/3 on beauty baryon polarization. This file also
includes three postscript figures (uuencoded), which 
are attached at the end.
  
Best regards,                  | phone:++42-2-6605-2159  
Hrivnac, Lednicky, Smizanska   | fax:++42-2-8218227
Institute of Physics           | Internet:
Na Slovance 2, Prague          | smizansk@hp10.fzu.cz
Czech Republic                 |

\documentstyle[12pt,epsfig,amstex]{article}
\makeatletter
\def\pranum#1{\gdef\@pranum{#1}}\gdef\@pranum{}
\def\prepranum#1{\gdef\@prepranum{#1}}
\def\@prepranum{PRA--HEP--} % initial logo
\def\pranumtwo#1{\gdef\@pranumtwo{#1}}\gdef\@pranumtwo{}
\def\praheader{%
\ifx\@pranum\@empty\vglue 5ex
\else
\halign to\textwidth{\hfil ## \hfil\tabskip 0in
plus \textwidth & \hfil ## \tabskip0pt\cr
{\sl Institute of Physics, Acad.\ of Sci.\ of the Czech Rep.} &
{\@prepranum\@pranum}\cr
{\sl and} & \ifx\@pranumtwo\@empty\@date\else\@pranumtwo\fi\cr
{\sl Nuclear Centre, Charles University} & 
\ifx\@pranumtwo\@empty\else\@date\fi\cr
{\sl Prague} & \cr}\fi}
\def\titlepage{\clearpage%
\thispagestyle{empty}\pagestyle{plain}\pagenumbering{arabic}%
\setcounter{footnote}{0}\setcounter{page}{0}%
\def\thefootnote{\fnsymbol{footnote}}
\let\footnotesize\small
\praheader\vskip8em}
\def\submitted#1{\vskip1em\begin{center}\it#1\end{center}}
\def\titlefont{\LARGE}
\def\title#1{%
\vskip1em\begin{center}\titlefont #1\end{center}\vskip1.5em}
\def\@makefnmark{\hbox{$^{\@thefnmark)}$}}
\def\author#1{%% Treat the list of authors
\begingroup
\global\@topnum\z@ \begin{center}{\lineskip.5em
#1}
\end{center}\par\@thanks\endgroup}
\def\address#1{% The institute name in italic
\begin{center}{\lineskip.5em\it #1}
\end{center}\par}
\def\abstract{\vskip3em\begin{center}{\bf \abstractname}\end{center}}

\def\abstractname{Abstract}
\def\endtitlepage{%% Reset counters
\def\thefootnote{\arabic{footnote}}
\setcounter{footnote}{0}\let\titlepage\relax\vfill
\newpage\setcounter{page}{1}\pagestyle{plain}\pagenumbering{arabic}%
\gdef\@thanks{}\gdef\@author{}\gdef\@title{}\gdef\@pranum{}
\gdef\@prepranum{}\gdef\@pranumtwo{}
\let\thanks\relax\let\praheader\relax}
\makeatother
\font\lowrm=cmr10 scaled \magstep 0

\newcommand{\Lamzero}{\mbox{$\Lambda^{0}\ $}}
\newcommand{\Lamzerob}{\mbox{$\Lambda^{0}_{b}\ $}}
\newcommand{\Lamzeroto}{\mbox{$\Lamzero \rightarrow p \pi^{-}\ $}}
\newcommand{\Lamzerobto}{\mbox{$\Lamzerob \rightarrow \LamzeroJpsi\ $}}

\newcommand{\aLamzero}{\mbox{$\overline{\Lambda^{0}}\ $}}
\newcommand{\LamzeroJpsi}{\mbox{$\Lamzero J/\psi\ $}}
\newcommand{\LamzeroJpsitol}{\mbox{$\LamzeroJpsi \rightarrow
 p\pi^{-}\l^{+}\l^{-}\ $}}
\newcommand{\LamzeroJpsitomu}{\mbox{$\LamzeroJpsi \rightarrow
 p\pi^{-}\mu^{+}\mu^{-}\ $}}
\newcommand{\LamzeroJpsitoe}{\mbox{$\LamzeroJpsi \rightarrow p\pi^{-}
 e^{+}e^{-}\ $}}

\newcommand{\Xizerob}{\mbox{$\Xi^{0}_{b}\ $}}

\newcommand{\Xizerobto}{\mbox{$\Xizerob \rightarrow \LamzeroJpsi\ $}}
\newcommand{\Xinulabto}{\mbox{$\Xizerob \rightarrow \Xi^{0} J/\psi\ $}}

\newcommand{\Xizeroto}{\mbox{$\Xi^{0} \rightarrow \Lamzero \pi^{0}\ $}}
\newcommand{\Ximinusbto}{\mbox{$\Xi^{-}_{b} \rightarrow \Xi^{-} J/\psi\ $}}
\newcommand{\Ximinusto}{\mbox{$\Xi^{-} \rightarrow \Lamzero \pi^{-}\ $}}

\newcommand{\Bto}{\mbox{$B_{b} \rightarrow \LamzeroJpsi\ $}}
\newcommand{\Jpsitomm}{\mbox{$J/\psi \rightarrow \mu^{+} \mu^{-}\ $}}
\newcommand{\Jpsitoee}{\mbox{$J/\psi \rightarrow e^{+} e^{-}\ $}}
\newcommand{\Jpsitol}{\mbox{$J/\psi \rightarrow l^{+} l^{-}\ $}}
\newcommand{\mumuppi}{\mbox{$\mu^{+}\mu^{-}p\pi^{-}\ $}}
\newcommand{\mueeppi}{\mbox{$\mu eep\pi^{-}\ $}}
\newcommand{\bquark}{\mbox{$\rm b\ $}}
\newcommand{\abquark}{\mbox{$\rm \overline{b}\ $}}
\newcommand{\bbarb}{\mbox{$\rm pp \rightarrow \bquark \abquark X\ $}}

\newcommand{\jnn}{\mbox{${1 \over N}$}}
\newcommand{\jnd}{\mbox{${1 \over 2}$}}
\newcommand{\jnt}{\mbox{${1 \over 3}$}}
\newcommand{\jnp}{\mbox{${1 \over 5}$}}
\newcommand{\jnde}{\mbox{${1 \over 9}$}}
\newcommand{\jnpt}{\mbox{${1 \over 15}$}}
\newcommand{\jnsp}{\mbox{${1 \over 45}$}}
\newcommand{\snstp}{\mbox{${16 \over 135}$}}
\newcommand{\dnstp}{\mbox{${2 \over 135}$}}
\newcommand{\dnsp}{\mbox{${2 \over 45}$}}

\newcommand{\jnst}{\mbox{${1 \over {(4 \pi)}^{3}}$}}
\newcommand{\tnd}{\mbox{${3 \over 2}$}}
\newcommand{\tnsd}{\mbox{${3 \over\sqrt{2}}$}}

\newcommand{\psubt}{\mbox{$p_{t}\ $}}
\newcommand{\obr}[4]{
  \begin{figure}[h]
    \begin{center}
      \mbox{\epsfig{file=#1,height=#4}}
      \end{center}
    \caption{#2}
    \label{#3}
  \end{figure}}

\newcommand{\tabulka}[3]{
  \begin{table}[h]
    \begin{center}
      #1
      \end{center}
    \caption{#2}
    \label{#3}
  \end{table}}

\pranum{94/3}
\pranumtwo{hep-ph/9405231}
\date{May 5, 1994}
\begin{document}

\begin{titlepage}

  \title{Feasibility of Beauty Baryon Polarization Measurement in
         \LamzeroJpsi decay channel by ATLAS-LHC}

  \author{Julius H\v{r}ivn\'{a}\v{c},
          Richard Lednick\'{y} and
          M\'{a}ria Smi\v{z}ansk\'{a}}

  \address{\lowrm Institute of Physics AS CR\\
               Prague, Czech Republic}

  \submitted{submitted to Zeitschrift f\"{u}r Physik C}

  \date{}

  \begin{abstract}

    The possibility of beauty baryon polarization measurement by cascade decay
angular distribution
    analysis in the channel \LamzeroJpsitol is demonstrated.
    The error analysis shows that in the proposed LHC experiment ATLAS at the
luminosity
    $10^{4} pb^{-1}$ the polarization can be measured with the statistical
precision
    better than
    $\delta=0.010$ for \Lamzerob and $\delta=0.17$ for \Xizerob.

    \end{abstract}

  \end{titlepage}

\section*{Introduction}
   The study of polarization effects in multiparticle production provides an
 important information
   on spin-dependence of the quark confinement. Thus substantial polarization
of
 the hyperons produced in nucleon
   fragmentation processes \cite{Heller,Pondrom} as well as the data on the
 hadron polarization asymmetry
   were qualitatively described by recombination quark models taking into
 account the leading effect
%   due to the valence hadron constituents \cite{Andersson} $-$
% \cite{Lednicky1}.
    due to the valence hadron constituents $[3-6]$.
   Although these models correctly predict  practically zero polarization of
 $\overline{p}$,$\overline{\Lambda}$
   and $\Omega^{-}$, they fail to explain the large polarization   of
   antihyperons $\overline{\Xi^{+}}$ and $\overline{\Sigma^{-}}$ recently
 discovered in Fermilab \cite{Ho,Morelos}.

   The problem of quark polarization effects
 could be clarified in polarization
   measurements involving  heavy quarks. In particular, an information about
the
 quark mass dependence of these effects
   could be obtained \cite{Grant,Lednicky2}. The polarization is expected to be
   proportional to the quark mass if it arises due to scattering on a colour
% charge\cite{Kane,Szwed}. The opposite
charge $[10-12]$. The opposite
   dependence takes place if the quark becomes polarized due to the interaction
 with an ''external'' confining field,
    e.g.,
   due to the effect of spontaneous radiation polarization \cite{Batyunya}. The
 decrease of the polarization with
   increasing quark mass is expected also in the model of ref. \cite{Troshin}.

   In QCD the polarization might be expected to vanish with the quark mass due
 to vector character of the quark-gluon
   coupling \cite{Kane}. It was shown however in Ref.\cite{Efremov} that the
 quark mass should be effectively
   replaced by the hadron mass M so that even the polarization of ordinary
 hadrons can be large. The polarization
   is predicted to be independent of energy and to vanish in the limit of both
 low and high hadron transverse
   momentum \psubt. The maximal polarization $P_{max}(x_{F})$ is reached at
 $\psubt \approx M$ and
  depends on the Feynman variable $x_{F}$. Its magnitude (and in particular its
 mass dependence) is determined by two
  quark-gluon correlators which are not predicted by perturbative QCD.

   The  polarization of charm baryons in hadronic reactions is still unmeasured
 due to the lack of sufficient  statistics.
   Only some indications on a nonzero polarization were reported
\cite{BIS,ISR}.
   For beauty physics the future experiments on LHC or HERA give an opportunity
 to obtain large statistical samples
    of beauty baryon ($\Lamzerob ,\Xizerob $) decays into \LamzeroJpsitol,
    which is favorable mode to detect experimentally. Dedicated triggers for
 CP-violation effects in $b$-decays,
    like the  high-\psubt one-muon trigger (LHC) \cite{ATLAS} or the $J/\psi $
 trigger (HERA) \cite{HERA}
     are selective also for this channel.

    Below we consider the possibility of polarization measurement of  beauty
 baryons \Lamzerob and \Xizerob
     with the help of cascade decay
    angular distributions in the channel $\Lamzerob(\Xizerob) \rightarrow
 \LamzeroJpsitol$.

    \section*{Polarization measurement method and an estimation of the
 statistical error.}

   In the case of parity nonconserving  beauty baryon ($B_{b}$) decay the
 polarization causes the asymmetry
   of the distribution of the cosine of the angel $\theta$
   between the beauty baryon decay and production analyzers:
   \begin{equation}\label{eq:w0}
     w(\cos\theta) = {1\over2} (1 + \alpha_{b} P_{b} \cos\theta),
     \end{equation}
   where $P_{b}$ is a projection of the polarization vector on the production
   analyzer and $\alpha_{b}$ is a decay asymmetry parameter. As the
polarization
 vector
   is perpendicular to  $B_{b}$ production plane  (due to parity conservation
   in the production process),
   the production analyzer
   should be directed parallel to the normal to this plane: $\vec{n} =  {
 \vec{p}_{inc} \times \vec{p}_{B_{b}}
             \over  |\vec{p}_{inc} \times \vec{p}_{B_{b}}|}$, where
 $\vec{p}_{inc}$ and $\vec{p}_{B_{b}}$ are momenta
    of incident particle and $B_{b}$ in c.m. system.

    The asymmetry parameter
   $\alpha_{b}$ characterizes parity nonconservation in
   a weak decay of $B_{b}$  and depends on the choice
   of the decay analyzer. In the two-body decay \Bto it is natural to choose
   this  analyzer oriented in the direction of
   \Lamzero momentum $\vec{p}_{\Lamzero}$ in the $B_{b}$ rest system. The
 considered decay  can be described
   by 4  helicity amplitudes $A(\lambda_{1},\lambda_{2})$ normalized to unity:
   $a_{+}=A(1/2,0)$,$a_ {-}=A(-1/2,0)$,
   $b_{+}=A(-1/2,1)$ and $b_{-}=A(1/2,-1)$,

   \begin{equation}\label{eq:no}
    \vert  a_{+}\vert ^2 + \vert  a_{-}\vert ^2 + \vert
    b_{+}\vert ^2 + \vert  b_{-}\vert ^2=1.
    \end{equation}
    The difference of \Lamzero and $J/\psi$
   helicities $\lambda_{1}$ - $\lambda_{2}$
   is just a projection of $B_{b}$ spin onto the decay analyzer.
   The decay asymmetry parameter $\alpha_{b}$ is expressed through these
 amplitudes in the form

   \begin{equation}\label{eq:alfa}
   \alpha_{b}=\vert  a_{+}\vert ^2 - \vert  a_{-}\vert ^2 + \vert  b_{+}\vert
^2
 - \vert  b_{-}\vert ^2.
   \end{equation}
   If P-parity in $B_{b}$
   decay were conserved, then $\vert  a_{+}\vert ^2 = \vert  a_{-}\vert ^2$,
    $\vert  b_{+}\vert ^2 = \vert  b_{-}\vert ^2$ so that $\alpha_{b}$ would be
 0.
    In the case of known and sufficiently nonzero value of $\alpha_{b}$ the
 beauty baryon polarization
    could be simply measured with the help of angular distribution
(\ref{eq:w0})
 (see, e.g., \cite{FRIDMAN}).
    Due to lack of experimental information and rather uncertain theoretical
 estimates of $\alpha_{b}$
    for the decay \Lamzerobto \cite{BALL} both the polarization and
$\alpha_{b}$
 (or the decay amplitudes)
    should be determined simultaneously. This can be achieved with the help of
 information on
    \Lamzero and $J/\psi$ decays. Though it complicates the analysis, it should
 be stressed that the measurement
     of the beauty baryon decay amplitudes could give valuable constrains on
 various theoretical models. Generally,
     such a measurement can be done provided that
     at least one of the secondary decays is asymmetric and its decay asymmetry
 parameter
   is known \cite{Lednicky2}. In our case it is the decay \Lamzeroto with the
 asymmetry parameter
    $\alpha_{\Lambda}$=0.642.

    The angular distribution
   in the cascade decay $B_{b} \rightarrow \LamzeroJpsitol$ follows directly
 from Eq. $(6)$ of \cite{Lednicky2},
   taking into account that the only nonzero multipole parameters in the decay
 \Jpsitol
   are $T_{00}$=1 and $T_{20}$=${1 \over\sqrt{10}}$.
   It can be written in the form

    \begin{equation}\label{eq:w}
%      \begin{split}
       w(\Omega ,\Omega_{1}, \Omega_{2}) =  \jnst
       \sum_{i=0}^{i=19} f_{1i} f_{2i}(P_{b} ,\alpha_{\Lambda}) F_{i}(\theta,
       \theta_{1},\theta_{2},\phi_{1},\phi_{2}),
%        \end{split}
      \end{equation}
    where $f_{1i}$ are bilinear combinations of the decay amplitudes
    $a_{+}$, $a_{-}$, $b_{+}$, $b_{-},$
    $f_{2i}(P_{b},\alpha_{\Lambda})$ stands for $P_{b}\alpha_{\Lambda}$,
 $P_{b}$, $\alpha_{\Lambda}$
    or 1 and $F_{i}$ are orthogonal angular functions (Table\ref{table:abab}).
    $\Omega$=($\theta$,$\phi$)  are the polar and the azimuthal angles of
     the \Lamzero momentum $\vec{p}_{\Lamzero}$  in the
    $B_{b}$ rest frame  with z-axis parallel to the production normal $\vec{n}$
 ( the choice of x and y axes
    is not important since parity conservation in the production process
 guarantees independence
    of the azimuthal angle $\phi$),
    $\Omega_{1}$=($\theta_{1}$,$\phi_{1}$) are the angles of proton momentum
     in the \Lamzero rest frame with axes defined as
    $z_{1}\uparrow\uparrow\vec{p}_{\Lamzero}$,
    $y_{1}\uparrow\uparrow\vec{n}\times\vec{p}_{\Lamzero}$,
    $\Omega_{2}$=($\theta_{2}$,$\phi_{2}$) are the  angles of the momentum of
    one of the decay leptons in the $J/\psi$ rest frame   with axes defined as
    $z_{2}\uparrow\uparrow\vec{p}_{J/\psi}$,
    $y_{2}\uparrow\uparrow\vec{n}\times\vec{p}_{J/\psi}$
 ($\vec{p}_{J/\psi}$=$-\vec{p}_{\Lamzero}$
    is  ${J/\psi}$ momentum in the $B_{b}$ rest frame).

    This distribution depends on 7 unknown independent parameters. One of them
 is the polarization
    $P_{b}$ and six others determine the four complex amplitudes
    \begin{equation}
%     \begin{split}
     a_{+}=\vert  a_{+}\vert  e^{i\alpha_{+}},
     a_{-}=\vert  a_{-}\vert  e^{i\alpha_{-}},
     b_{+}=\vert  b_{+}\vert  e^{i\beta_{+}},
     b_{-}=\vert  b_{-}\vert  e^{i\beta_{-}}.
%      \end{split}
      \end{equation}
    Taking into account the normalization condition and arbitrariness of the
 common phase, these
    parameters can be chosen as

    \begin{equation}\label{eq:pa}
%     \begin{split}
      \alpha_{b},
      r_{0}= \vert  a_{+}\vert ^2 + \vert  a_{-}\vert ^2,
    \vert  a_{+}\vert ^2 - \vert  a_{-}\vert ^2,\alpha_{+},\alpha_{-},
    \chi = \alpha_{+} + \alpha_{-} - \beta_{+} - \beta_{-}.
%    \end{split}
      \end{equation}
       The parameters can be determined from
    the measured decay angles by a 5-dimensional likelihood fit or by the
moment
 method:
      the angular distribution of the form  (\ref{eq:w}) allows one to
introduce
 19 moments
    $<F_{i}> \sim f_{1i}.f_{2i}$ and determine the parameters from their
 measured values by weighted least-squares method.
       For the polarization measurement the formula (\ref{eq:w})
    integrated over the azimuthal angles
    $\phi_{1}$, $\phi_{2}$ would be in principle sufficient \cite{Lednicky2}.
In
 this case the
    number of free parameters is reduced to 4 ( the phases don't enter) and
only
 a 3-dimensional
    fit is required. We will see, however, that
    the information on these angles may  substantially increase the
    precision of the $P_{b}$ determination.

       To simplify the error analysis, we follow ref. \cite{Lednicky2} and
 consider here only the most
    unfavourable situation, when the parameters $P_{b}^2$,$\vert  a_{+}\vert ^2
 - \vert  a_{-}\vert ^2$ and
     $\vert  b_{+}\vert ^2 - \vert  b_{-}\vert ^2$ are much smaller than
 $\alpha_{\Lambda}^2$. In this case
     the moments $<F_{i}>$ can be considered to be independent, having the
 diagonal error matrix
    \begin{equation}
%     \begin{split}
     W=\jnn  diag( \jnt, \jnt, \jnde, \jnp, \jnpt, \jnpt, \jnsp,
      \snstp, \snstp,
                          \snstp, \snstp, \dnstp, \dnstp, \dnstp, \dnstp,
\dnsp,
 \dnsp,
                            \dnsp, \dnsp ).
%     \end{split}
      \end{equation}

     Here $N$ is a number of $B_{b}$ events (assuming that the background can
be
 neglected, see next section).
     The error matrix $V$ of the vector $\vec{a}$ of the parameters
 ${a_{j}}$,$j$=1,..7 defined in (\ref{eq:pa}) is
     \begin{equation}
%     \begin{split}
     V(\vec{a})=(A^{T} W^{-1} A)^{-1},
%     \end{split}
      \end{equation}
     where the elements of the matrix $A$ are $A_{ij}={d(f_{1i}.f_{2i}) \over
 da_{j}}$.
     In the considered situation  the error on the polarization $P_{b}$ is
      \begin{equation}\label{eq:chy}
%     \begin{split}
     \delta \equiv \sqrt {V_{11}} = {\delta _{0} \over \sqrt {N}},
%     \end{split}
      \end{equation}
      \begin{equation}\label{eq:chy1}
%      \begin{split}
     \delta _{0}= {1\over
     \sqrt{
     \alpha_{\Lambda}^{2} . [{(2 r_{0}-1)^{2}\over {9}} + {(r_{0}+1)^{2}\over
 180}
     +{4 r_{0}^{2}\over {15}} + {(1-r_{0})^{2}\over {15}}+ {(1-r_{0})
 (1+\cosh\chi) \over{15}}]+
     {r_{0} (1-r_{0})(1-\cosh\chi) \over{10}}
           }}.
%       \end{split}
      \end{equation}

    Here $\delta _{0}$  depends only on the
    relative contribution $r_{0}$ of the decay amplitudes with helicity
 $\lambda_{2} = 0$
     and on the relative phase $\chi$ (Figs. \ref{obr:polariz}a,b).
    The maximal error on $P_{b}$ is
    $\delta_{\max}={4.7\over\sqrt{N}}$ and it
    corresponds to the case when
    $r_{0}={1 \over 3}$ and the phase $\chi = 0$. The minimal error
    in the considered most unfavorable situation is
    $\delta_{\min}={2.5\over\sqrt{N}}$.
    It is obtained in the case when only $\lambda_{2} = 0$ helicity amplitudes
    contribute ($r_{0}=1$), independently of their phases. If the information
    on the azimuthal angles would be neglected  then
 $\delta_{\max}={14.0\over\sqrt{N}}$
    and $\delta_{\min}={4.3\over\sqrt{N}}$.

  \section*{Estimates of indirect  \Lamzerob
    and \Xizerob  production, background processes and possible statistics at
 ATLAS-LHC  }

 The beauty baryons  \Lamzerob and \Xizerob are produced directly or through
the
 decays of
 heavier states. According to the PYTHIA tables we consider here the strong
 decays
 $\Sigma_{b} \rightarrow \Lamzerob \pi$ , $\Sigma_{b}^{*} \rightarrow
\Lamzerob
 \pi$
 and the electromagnetic decays $\Xi^{0'}_{b}\rightarrow \Xizerob \gamma$ or
 $\Xi^{0*}_{b}\rightarrow \Xizerob \gamma$.The observable polarization
$P_{obs}$
 depends on the polarizations $P_{B_{b}}$  of direct beauty baryons  and their
  production fractions $b_{B_{b}}$ (i.e. probabilities of the b-quark to
 hadronize to  certain
  baryons $B_{b}$). In considered decays the beauty baryon \Lamzerob
  or \Xizerob retains ${-1\over3} ({1\over3})$ of the polarization of a parent
  with spin ${1\over2}^{+} ({3\over2}^{+})$ (see Appendix).
 For $P_{obs}$  we get:

\begin{equation}\label{eq:pobs1}
%     \begin{split}
 P_{obs}={
          b_{\Lambda^{0}_{b}}P_{\Lambda^{0}_{b}}+\sum_{i}({-1\over3}
 b_{\Sigma_{bi}}P_{\Sigma_{bi}}
                                    +{1\over3} b_{\Sigma_{bi}^{*}}
 P_{\Sigma_{bi}^{*}})
          \over
          b_{\Lambda^{0}_{b}}+\sum_{i}(b_{\Sigma_{bi}}+b_{\Sigma_{bi}^{*}})
           },
%      \end{split}
      \end{equation}
for \Lamzerob and for \Xizerob:
\begin{equation}\label{eq:pobs2}
%     \begin{split}
 P_{obs}={
          b_{\Xi^{0}_{b}}P_{\Xi^{0}_{b}}+{-1\over3}
 b_{\Xi^{0'}_{b}}P_{\Xi^{0'}_{b}}
                                +{1\over3} b_{\Xi^{0*}_{b}}P_{\Xi^{0*}_{b}}
          \over
          b_{\Xi^{0}_{b}}+b_{\Xi^{0'}_{b}} + b_{\Xi^{0*}_{b}}
          }.
%      \end{split}
      \end{equation}
 The  summation goes over positive, negative and neutral $\Sigma_{b}$ and
$\Sigma_{b}^{*}$.
Assuming the polarization of the heavier states to be similar in magnitude to
 that
of directly produced \Lamzerob or \Xizerob ($P_{\Lambda^{0}_{b}}$ or
 $P_{\Xi^{0}_{b}}$)
we may expect the observed polarization in an interval of $(0.34-0.67)
 P_{\Lambda^{0}_{b}}$
for \Lamzerob and $(0.69-0.84) P_{\Xi^{0}_{b}}$ for \Xizerob.

  The polarization can be measured for \Lamzerob and \Xizerob baryons
  and for their antiparticles. \Lamzerob (\Xizerob) are unambigously
  distinquishable from their antiparticles by effective mass of $p \pi^{-}$
 system from
  \Lamzeroto decay.  The wrong assignment of antiproton and pion masses gives
 the kinematical
%  reflection of \Lamzero starting   at $1.5\ GeV$ (Fig. \ref{obr:lambda})
%%which
% is far from
  reflection of \aLamzero starting   at $1.5\ GeV$  which is far from
  \Lamzero mass comparing to the  mass resolution of \Lamzero produced in  the
 decays
  of \Lamzerob or  \Xizerob.

  \Lamzerob and \Xizerob can be distinguished from each other by the effective
 mass of the
  $p\pi^{-}\l^{+}\l^{-}$ system.
  The mass difference of \Lamzerob and \Xizerob is $220MeV$. The
  mass resolution in this region (taking into account the table mass values for
  \Lamzero and $J/\psi$ candidates) is $\sigma_{\Lambda^{0} J/\psi}\approx
 26MeV$ for \LamzeroJpsitomu \\
channel. Thus even  the background from
  \Lamzerob in the region of masses $m_{\Xi_{b}^{0}}\pm 3\sigma_{\Lambda^{0}
 J/\psi}$ is negligible.

  There are several other sources of background to \LamzeroJpsi decays of
 \Lamzerob and \Xizerob.
    The dominant background  comes from  $J/\psi$ from b-hadron
  decays and \Lamzero produced in fragmentation or in  the decay of the same
 b-hadron.
  After the cutting off the  low transverse momenta (less than $0.5 GeV$) of
 $p$ and $\pi$ from \Lamzero decay
  this background can be reduced to $\approx 1.5\%$ in a region of
   $m_{\Lambda^{0}_{b}} \pm 3\sigma_{\Lambda^{0}_{b} J/\psi}$(Figs.
 \ref{obr:bckla}a,b).
  The remaining background comes mainly from two processes: \Xinulabto ,
 \Xizeroto and \Ximinusbto ,
  \Ximinusto. For  \Xizerob this  background (Fig.\ref{obr:bckksi}a ) is more
 important ( $\approx 20\%$). The reason is that
   the decay \Xizerob is governed  by $b\rightarrow d c\overline{c}$ process
 which is
  Cabbibo suppressed by a factor ${\sin (\theta_{c})}^2=0.22^{2}$ compared to
 the decays \Xinulabto,
  \Ximinusbto and also \Lamzerobto, which are governed by  $b\rightarrow s
 c\overline{c}$. However
   \Lamzero from \Xinulabto or \Ximinusbto\\
    is produced in a weak hyperon
 decay, so  this background
  can be efficiently reduced by the cut on the minimal distance $d$ between
 $J/\psi$ and \Lamzero.
  A conservative cut $d<1.5 mm$ reduces this background  by a factor $\approx
 0.05$ (Fig. \ref{obr:bckksi}b).

     The background from $B^{0}_{d}\rightarrow J/\psi K^{0}$ when one of $\pi$
  mesons is considered as a proton is negligible after
  the effective mass cuts
%  on ($p \pi$) and ($p \pi J/\psi$) systems (Fig.\ref{obr:bck_bo_mumu}).
  on ($p \pi$) and ($p \pi J/\psi$) systems.

    Background from fake $J/\psi's$, as it has been shown in  \cite{ATLAS}, can
 be reduced to a low level
    by cuts on the distance between the primary vertex and the production point
 of the $J/\psi$ candidate
    and the distance of closest approach between
    the two particles from the decay. These cuts also suppress the background
 from real $J/\psi's$
    comming directly from hadronization.

  The number of produced \Lamzerob and \Xizerob is calculated for the cross
 section of
  \bbarb \\
  equal to $500 \mu b$.
  The  production fraction $b_{\Lambda^{0}_{b}}$ of  $b\rightarrow\Lambda_{b}$
  multiplied by branching ratio $br_{\Lambda^{0}_{b}}$ of \Lamzerobto decay
was
 measured
  in two experiments: UA1 \cite{UA1} gives the value
  $b_{\Lambda^{0}_{b}}.br_{\Lambda^{0}_{b}}={1.8 \pm 0.6}\ 10^{-3}$ while CDF
 \cite{CDF1} put only an upper limit
  on this value of $0.5\ 10^{-3}$ which we'll use as a conservative
  estimate in statistics and error calculations.
  From PYTHIA generator $b_{\Lambda^{0}_{b}}=0.08$ including also \Lamzerob
 produced in strong $\Sigma_{b}$
  decays. If we neglect possible changes of the production fractions with the
 energy
   the value of $br_{\Lambda^{0}_{b}}=2.2\ 10^{-2}$  can be derived from UA1
 result or the upper limit value on
  $br_{\Lambda^{0}_{b}}$ of $0.6\ 10^{-2}$  from CDF result.
  The production fraction of  \Xizerob given by PYTHIA is $b_{\Xi^{0}}=5.5\
 10^{-3}$,  where also
  \Xizerob from   electromagnetic decays of $\Xi^{0'}_{b}$ and $\Xi^{0*}_{b}$
 are taken into account.
  The branching ratio of the decay \Xizerobto can be roughly estimated
 multiplying $br_{\Lambda^{0}_{b}}$
  by the Cabbibo suppression factor ${\sin (\theta_{c})}^2=0.22^{2}$:
 $br_{\Xi^{0}_{b}}=1.1\ 10^{-3}$
  (or $br_{\Xi^{0}_{b}}<0.3\ 10^{-3}$) using UA1 (CDF) results.
  Thus the number of \Lamzerobto
  processes  can be expected to be $\approx 300$ times larger than the number
of
  \Xizerobto ones.

  Both channels \LamzeroJpsitomu and \LamzeroJpsitoe can be used
  for the analysis. However statistics (Table\ref{table:results}) will be
 dominated by the
  former decay as the trigger
  can be satisfied
  by one of the two  muons. In the case of \LamzeroJpsitoe the trigger comes
 only from
  semileptonic decay  $b\rightarrow\mu X$ of the associated b-hadron.
  The reconstruction efficiencies are calculated by demanding that the events
 satisfy the cuts listed bellow.
  The first set of cuts corresponds the requirements set by the trigger:

  - For \LamzeroJpsitomu decay one muon is required to satisfy the single muon
 trigger conditions:
  $p^{\mu}_{\perp}>6GeV$ and $|\eta|<1.6$.  For the second muon the cuts
 $p^{\mu}_{\perp}>3GeV$,$|\eta|<2.5$ are applied,
   supposing that within these values muon  can be identified using the last
 segment of the hadron calorimeter by its minimum
   ionizing signature.

   - For \LamzeroJpsitoe decay both electrons are required to have
 $p^{e}_{\perp}>1GeV$. The low threshold for
   electrons is possible, because of electron identification in the transition
 radiation tracker (TRT) \cite{Gavrilenko}.
   The events are required to contain one muon with a $p^{\mu}_{\perp}>6GeV$
and
 $|\eta|<1.6$

  The second set of cuts corresponds to 'offline' analysis cuts. The same cuts
 as for $B^{0}_{d}\rightarrow J/\psi K^{0}$
  reconstruction \cite{ATLAS} can be used (the only exception is the mass
 requirement for \Lamzero candidate, see the last
   of the next cuts) :

  - The two charged hadrons from \Lamzero decay are required to be within the
 tracking volume $|\eta|<2.5$, and transverse
  momenta of both to be greater than $0.5GeV$.

  - \Lamzero decay length in the transverse plane with respect to the beam axis
 was required to be greater than $1cm$ and
  less than $50cm$. The upper limit ensures that the charged tracks from
 \Lamzero decay start before the inner radius of
   TRT, and that there is a space point from the innermost layer of the outer
 silicon tracker. The lower limit reduces
   the combinatorial background  from particles originating from the primary
 vertex.

  - The distance of closest approach between the two muon (electron) candidates
 forming the $J/\psi$ was required to be less
  than $320\mu m$ ($450 \mu m$), giving an acceptance for signal of 0.94.

  - The proper time of the \Lamzerob decay, measured from the distance between
 the primary vertex  and the
  production point of the $J/\psi$ in the transverse plane and the
reconstructed
 $p_{\perp}$ of \Lamzerob, was required
  to be greater than $0.5ps$. This cut is used to reduce the combinatorial
background, giving
 the acceptance for signal events $0.68$.

  - The reconstructed \Lamzero and $J/\psi$ masses were required to be within
 two standart deviations of nominal values.

  The results on expected  \Lamzerob and \Xizerob statistics and the errors of
their
 polarization measurement are summarized in
  Table\ref{table:results}.
    For both channels the statistics of reconstructed events at the luminosity
  $10^{4} pb^{-1}$ will be \\
  790 000 (220 000)  \Lamzerob and 2600 (720) \Xizerob,
 where the values are derived using
  UA1 (CDF)  results.

  For this statistics the maximal value of the statistical error on the
 polarization
  measurement, calculated from  formulae (\ref{eq:chy}) and (\ref{eq:chy1}),
 will be
  $0.005 (0.01)$ for \Lamzerob and $0.09 (0.17)$ for \Xizerob.

 \section*{Conclusion}

   At LHC luminosity $10^{4}pb^{-1}$ the beauty baryons
   \Lamzerob and  \Xizerob polarizations can be measured with the help of
 angular
%   distributions in the cascade decays \LamzeroJpsitomu and \LamzeroJpsitoe
%%with
   distributions in the cascade decays \LamzeroJpsitomu
    and \LamzeroJpsitoe with
 the statistical
   precision better than
  0.010 for \Lamzerob and 0.17 for \Xizerob.

  \appendix
 \section*{Appendix}

 The polarization transfered to \Lamzerob, which was produced indirectly in
 strong
    $\Sigma_{b}$ and $\Sigma_{b}^{*}$ decays, depends on the ratio ${\Delta
\over
 \Gamma}$
  of the mass difference $\Delta$ between
  $\Sigma_{b}$ and $\Sigma_{b}^{*}$ and the decay rate $\Gamma$ of
  $\Sigma_{b} \rightarrow  \Lamzerob \pi$ or
  $\Sigma_{b}^{*} \rightarrow  \Lamzerob \pi$ \cite{Falk}.
  We will consider the limit $\Delta >> \Gamma$ when the baryons
  $\Sigma_{b}$ and $\Sigma_{b}^{*}$ form two well separated resonances
  and their contributions to \Lamzerob polarization can be added incoherently.
  In such a case the polarization transfer can be easily calculated in a
 model-independent way
  using the angular distributions in  cascade decays \cite{Lednicky3}.
  Thus considering the cascade decay
  $\Sigma_{b} \rightarrow  \Lamzerob \pi$ or
  $\Sigma_{b}^{*} \rightarrow  \Lamzerob \pi$, \Lamzerobto, we have

   \begin{equation}\label{eq:wtr}
     w(\Omega,\Omega_{1}) \sim 1 +
     \alpha_{\Lambda^{0}_{b}} P_{\Sigma_{b}} ( \cos\theta \cos\theta_{1} \pm
     \sin\theta \sin\theta_{1} \cos\phi_{1} )
     \end{equation}
     for $\Sigma_{b}$ ( $\Sigma^{*}_{b}$).
Here  $\Omega$=($\theta$,$\phi$)  are the polar and the azimuthal angles of
the \Lamzerob momentum $\vec{p}_{\Lamzerob}$  in the
$\Sigma_{b} (\Sigma^{*}_{b})$ rest frame  with z-axis parallel to the
 production normal:
    $\vec{n} =  { \vec{p}_{inc} \times \vec{p}_{\Sigma_{b}}
             \over  |\vec{p}_{inc} \times \vec{p}_{\Sigma_{b}}|}$,
              where $\vec{p}_{inc}$ and $\vec{p}_{\Sigma_{b}}$ are momenta
    of incident particle and $\Sigma_{b}$ in c.m. system.
$\Omega_{1}$=($\theta_{1}$,$\phi_{1}$)  are the polar and the azimuthal
angles of \Lamzero in \Lamzerob rest frame with the axes defined as
    $z_{1}\uparrow\uparrow\vec{p}_{\Lamzerob}$,
    $y_{1}\uparrow\uparrow\vec{n}\times\vec{p}_{\Lamzerob}$.
After the transformation of  $\Omega_{1} \rightarrow \Omega_{1}^{'} $
 of \Lamzero angles  from the helicity frame $x_{1},y_{1},z_{1}$
 to the canonical frame $x,y,z$ with $z\uparrow\uparrow\vec{n}$
 and the integration over $\cos\theta$ and $\phi_{1}^{'}$
 we get the distribution of the cosine of the angle between the \Lamzero
 momentum vector (\Lamzerob decay analyzer) and the $\Sigma_{b} or
 \Sigma^{*}_{b}$
 production normal (which can be considered coinciding with the \Lamzerob
 production normal due to a small energy release in the $\Sigma_{b} or
 \Sigma^{*}_{b}$
 decays):
   \begin{equation}\label{eq:wtr2}
     w(\cos\theta^{'}_{1}) \sim 1 \mp
 {1\over3}    \alpha_{\Lambda^{0}_{b}} P_{\Sigma_{b}}  \cos\theta^{'}_{1}
     \end{equation}
 for $\Sigma_{b}$ ( $\Sigma^{*}_{b}$) decay.
 One can thus  conclude that \Lamzerob retains ${-1\over3} ({1\over3})$ of
  the $\Sigma_{b}$
 ($\Sigma^{*}_{b}$) polarization. The analogical analyses of electromagnetic
 decays  $\Xi^{0'}_{b}\rightarrow \Xizerob \gamma$ or
 $\Xi^{0*}_{b}\rightarrow \Xizerob \gamma$ show that \Xizerob retains
 ${-1\over3} ({1\over3})$ of the $\Xi^{0'}_{b} (\Xi^{0*}_{b})$ polarization.

\pagebreak

  \tabulka{\begin{tabular}{||c|c|c|c||}
             \hline\hline
              i & $f_{1i}$
           & $f_{2i}$                & $F_{i}$
                             \\
             \hline
              0 & $a_{+}a_{+}^{*}+a_{-}a_{-}^{*}+b_{+}b_{+}^{*}+b_{-}b_{-}^{*}$
           & 1                       & 1
                                 \\
             \hline
              1 & $a_{+}a_{+}^{*}-a_{-}a_{-}^{*}+b_{+}b_{+}^{*}-b_{-}b_{-}^{*}$
           & $P_{b}$                 & $\cos\theta$
                               \\
              2 & $a_{+}a_{+}^{*}-a_{-}a_{-}^{*}-b_{+}b_{+}^{*}+b_{-}b_{-}^{*}$
           & $\alpha_{\Lambda}$      & $\cos\theta_{1}$
                               \\
              3 & $a_{+}a_{+}^{*}+a_{-}a_{-}^{*}-b_{+}b_{+}^{*}-b_{-}b_{-}^{*}$
           & $P_{b}\alpha_{\Lambda}$ & $\cos\theta\cos\theta_{1}$
                             \\
              4 & $-a_{+}a_{+}^{*}-a_{-}a_{-}^{*}+\jnd b_{+}b_{+}^{*}+\jnd
 b_{-}b_{-}^{*}$ & 1                       & $d_{00}^{2}(\theta_{2})$
                                       \\
              5 & $-a_{+}a_{+}^{*}+a_{-}a_{-}^{*}+\jnd b_{+}b_{+}^{*}-\jnd
 b_{-}b_{-}^{*}$ & $P_{b}$                 & $d_{00}^{2}(\theta_{2})\cos\theta$
                                 \\
              6 & $-a_{+}a_{+}^{*}+a_{-}a_{-}^{*}-\jnd b_{+}b_{+}^{*}+\jnd
 b_{-}b_{-}^{*}$ & $\alpha_{\Lambda}$      &
 $d_{00}^{2}(\theta_{2})\cos\theta_{1}$                                  \\
              7 & $-a_{+}a_{+}^{*}-a_{-}a_{-}^{*}-\jnd b_{+}b_{+}^{*}-\jnd
 b_{-}b_{-}^{*}$ & $P_{b}\alpha_{\Lambda}$ &
 $d_{00}^{2}(\theta_{2})\cos\theta\cos\theta_{1}$                      \\
             \hline
              8 & $-3Re(a_{+}a_{-}^{*})$
 & $P_{b}\alpha_{\Lambda}$ &
 $\sin\theta\sin\theta_{1}\sin^{2}\theta_{2}\cos\phi_{1}$
 \\
              9 & $ 3Im(a_{+}a_{-}^{*})$
 & $P_{b}\alpha_{\Lambda}$ &
 $\sin\theta\sin\theta_{1}\sin^{2}\theta_{2}\sin\phi_{1}$
  \\
             10 & $-\tnd Re(b_{-}b_{+}^{*})$
 & $P_{b}\alpha_{\Lambda}$ &
 $\sin\theta\sin\theta_{1}\sin^{2}\theta_{2}cos(\phi_{1}+2\phi_{2})$
    \\
             11 & $ \tnd Im(b_{-}b_{+}^{*})$
 & $P_{b}\alpha_{\Lambda}$ &
 $\sin\theta\sin\theta_{1}\sin^{2}\theta_{2}sin(\phi_{1}+2\phi_{2})$
    \\
             \hline
             12 & $-\tnsd  Re(b_{-}a_{+}^{*}+a_{-}b_{+}^{*})$
 & $P_{b}\alpha_{\Lambda}$     &
 $\sin\theta\cos\theta_{1}\sin\theta_{2}\cos\theta_{2}\cos\phi_{2}$
       \\
             13 & $ \tnsd  Im(b_{-}a_{+}^{*}+a_{-}b_{+}^{*})$
 & $P_{b}\alpha_{\Lambda}$     &
 $\sin\theta\cos\theta_{1}\sin\theta_{2}\cos\theta_{2}\sin\phi_{2}$
        \\
             14 & $-\tnsd  Re(b_{-}a_{-}^{*}+a_{+}b_{+}^{*})$
 & $P_{b}\alpha_{\Lambda}$     &
 $\cos\theta\sin\theta_{1}\sin\theta_{2}\cos\theta_{2}\cos(\phi_{1}+\phi_{2})$
                   \\
             15 & $ \tnsd  Im(b_{-}a_{-}^{*}+a_{+}b_{+}^{*})$
 & $P_{b}\alpha_{\Lambda}$     &
 $\cos\theta\sin\theta_{1}\sin\theta_{2}\cos\theta_{2}\sin(\phi_{1}+\phi_{2})$
                   \\
             \hline
             16 & $ \tnsd  Re(a_{-}b_{+}^{*}-b_{-}a_{+}^{*})$            &
 $P_{b}$                  &
$\sin\theta\sin\theta_{2}\cos\theta_{2}\cos\phi_{2}$
                   \\
             17 & $-\tnsd  Im(a_{-}b_{+}^{*}-b_{-}a_{+}^{*})$            &
 $P_{b}$                  &
$\sin\theta\sin\theta_{2}\cos\theta_{2}\sin\phi_{2}$
                   \\
             18 & $ \tnsd  Re(b_{-}a_{-}^{*}-a_{+}b_{+}^{*})$            &
 $\alpha_{\Lambda}$       &
 $\sin\theta_{1}\sin\theta_{2}\cos\theta_{2}\cos(\phi_{1}+\phi_{2})$
       \\
             19 & $-\tnsd  Im(b_{-}a_{-}^{*}-a_{+}b_{+}^{*})$            &
 $\alpha_{\Lambda}$       &
 $\sin\theta_{1}\sin\theta_{2}\cos\theta_{2}\sin(\phi_{1}+\phi_{2})$
       \\
             \hline\hline
             \end{tabular}}
           {The coefficients $f_{1i}, f_{2i}$ and angular functions $F_{i}$ in
 distribution (\ref{eq:w}).}
           {table:abab}

  \tabulka{\begin{tabular}{||c||c|c|c||}
             \hline\hline
             Parameter                      & Value for \Lamzerob & Value for
 \Xizerob & Comment \\
             \hline\hline
             $L[cm^{-2}s^{-1}]$             & $10^{33}$           &      &
   \\
             \hline
             $t[s]$                         & $10^{7}$            &      &
   \\
             \hline
             $b(b\rightarrow B_{b})$        & $0.08$               & $5.5\
 10^{-3}$      &         \\
             \hline\hline
        $br(B_{b}\rightarrow\Lamzero J/\psi)$ & $2.2\ 10^{-2}$  & $1.1\ 10^{-3}
 $  &   \\
                                            & $(0.6\ 10^{-2})$  & $(0.3\
 10^{-3})$  &    \\
             \Jpsitomm                      & $0.06$              & $0.06$
&
         \\
             \Lamzeroto                     & $0.64$              & $0.64$
&
         \\
             \hline
             $br(b\rightarrow\mu X)
         br(B_{b}\rightarrow\Lamzero J/\psi)$ & $2.2\ 10^{-3}$  & $1.1\
10^{-4}$
    &    \\
                                            & $(0.6\ 10^{-2})$  & $(0.3\
 10^{-3})$  &    \\
             \Jpsitoee                      & $0.06$              & $0.06$
&
         \\
             \Lamzeroto                     & $0.64$              & $0.64$
&
         \\
             \hline\hline
             $\sigma(b\overline{b})$        & $500\mu b$          & $500\mu b$
    &         \\
             \hline\hline
             $N(\mumuppi)$ accepted         & $1535000$           & $5120$
 & $p^{\mu}_{\perp}                  >6GeV,|\eta|<1.6$ \\
                                            & $(426000)$          & $(1420)$
  & $p^{\mu}_{\perp}                  >3GeV,|\eta|<2.5$ \\
                                            &                     &      &
 $p^{\pi,p}_{\perp}                >0.5GeV,|\eta|<2.5$ \\
             \hline
             $N(\mueeppi)$ accepted         & $223000$            & $740$
&
 $p^{\mu}_{\perp}                  >6GeV,|\eta|<1.6$ \\
                                            & $(62000)$           & $(210)$
&
 $p^{e^{+},e^{-}}_{\perp}                >1GeV,|\eta|<2.5$ \\
                                            &                     &      &
 $p^{p,\pi^{-}}_{\perp}                 >0.5GeV,|\eta|<2.5$ \\
             \hline\hline
             $N(\mumuppi)$ reconstructed    & $720000$            & $2400$
 &         \\
                                            & $(200000)$          & $(670)$
 &         \\
             \hline
             $N(\mueeppi)$ reconstructed    & $65000$             & $220$
 &         \\
                                            & $(18000)$           & $(60)$
 &         \\
             \hline\hline
        the maximum statistical error       & $0.005$             & $0.09$
 &         \\
        on the polarization measurement     & $(0.010)$           & $(0.17)$
 &         \\
             $\delta(P_{b})$                &                     &
 &         \\
             \hline\hline
             \end{tabular}}
           {Summary on beauty baryon measurement possibilities for LHC
 experiment ATLAS.
           The values in brackets correspond to the CDF result, while the analogical
values
           without brackets  to the UA1 result.}
           {table:results}

 \obr{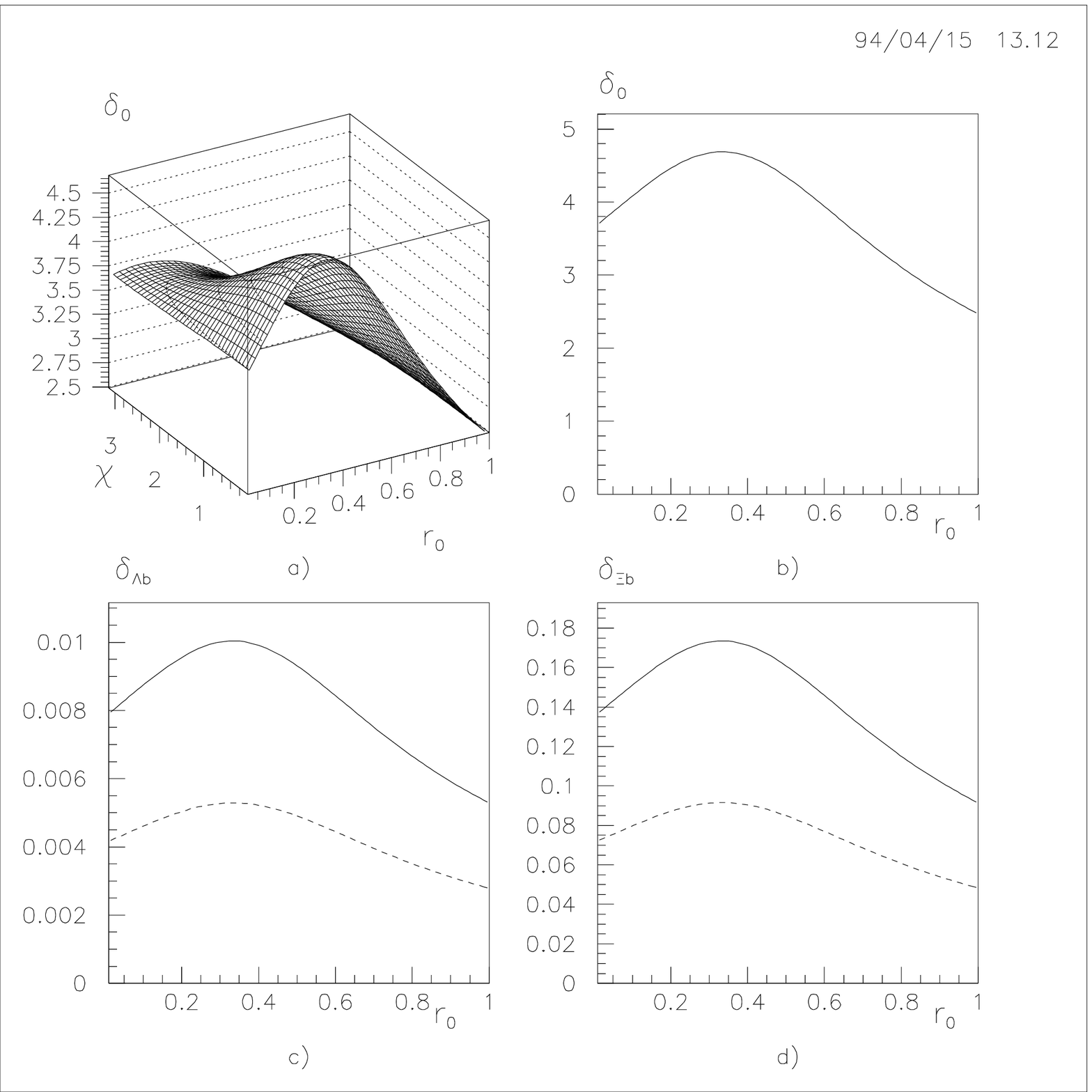}
      {\lowrm  The maximal statistical error on the polarization measurement
      $\delta(P_{b})$ and $\delta _{0}=N^{+{1\over2}}.\delta(P_{b})$ defined in
 (\ref{eq:chy1}):
       $\delta _{0}$ as a function of the relative contribution $r_{0}$ of the
 decay
       amplitudes with zero $J/\psi$ helicity
      and of the relative phase $\chi$ of the amplitudes with $J/\psi$ helicity
 equal to $0$ and $\pm 1 (a)$ .
       The  $\chi=0$ projection of this function (b) .
       Dependence of $\delta (P_{b})$ on $r_{0}$ at $\chi=0$ for \Lamzerob
 expected on Atlas
      at LHC luminosity
      $10^{4} pb^{-1}$  (c). Full (dotted) line corresponds to $\delta (P_{b})$
 derived from CDF (UA1) data.
       The same as (c) for \Xizerob (d).}
      {obr:polariz}
      {18cm}

%  \obr{lambda.eps}
%      {\lowrm \Lamzero/\aLamzero separation by effective mass spectrum of
% $p\pi^{-}$. Effective mass spectrum of \Lamzero from \Xizerobto (a).
%%Effective
% mass
% spectrum for the same $V_{0}$'s with proton mass assigned to negative track
%%and
% $\pi$ mass to positive track (b).}
%      {obr:lambda}
%      {18cm}

 \obr{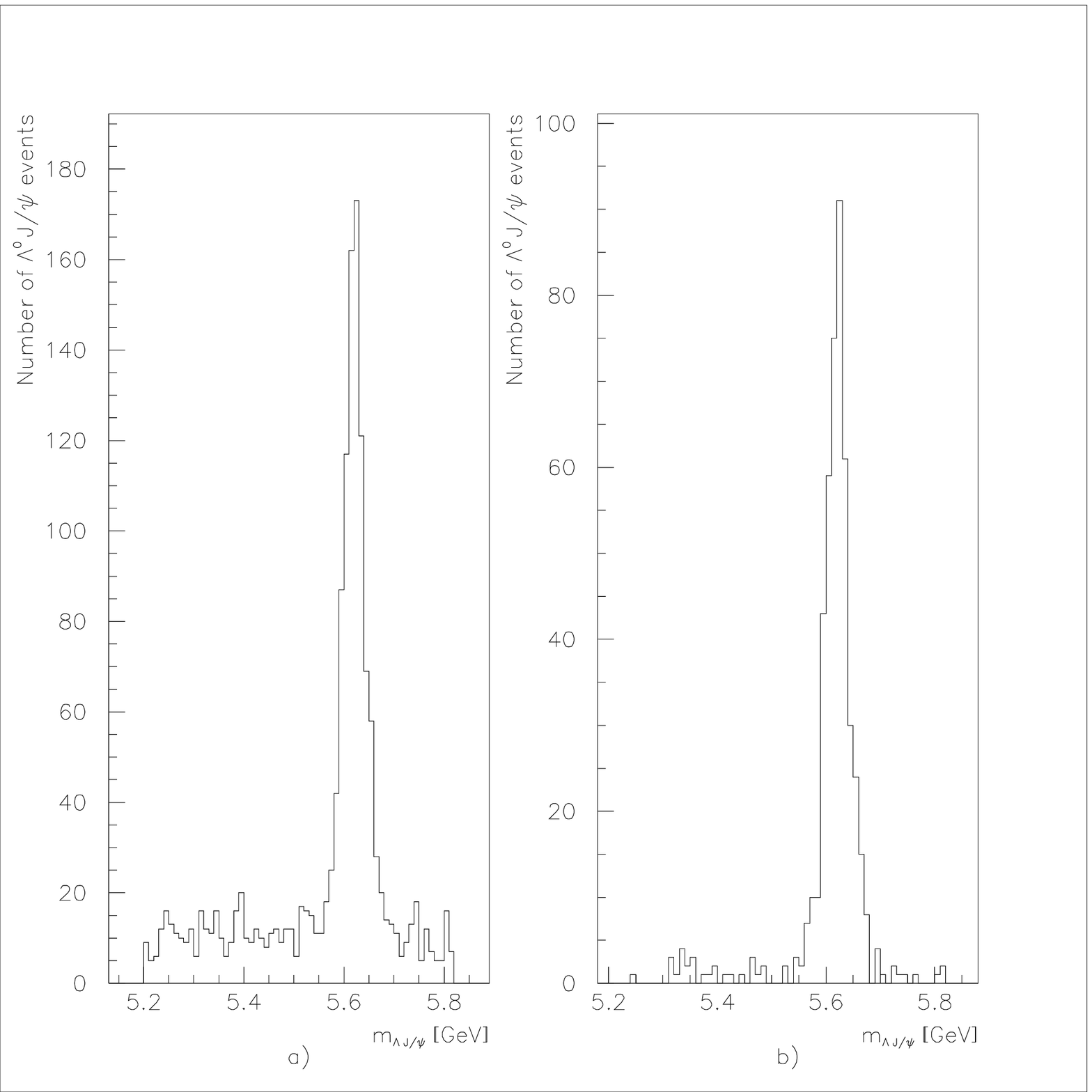}
      {\lowrm The \Lamzero $J/\psi$ effective mass distribution: The peak at
 $5.62 GeV$ is from
      \Lamzerob and background comes from $J/\psi$ from a b-hadron decay and
 \Lamzero either
      from the multiparticle production or from a b-hadron decay (a). The
events
 that passed
      the cut on the transverse momenta ($p_{T}>0.5 GeV$) for $p$ and $\pi^{-}$
 from \Lamzero decay
       (b).}
     {obr:bckla}
      {18cm}

 \obr{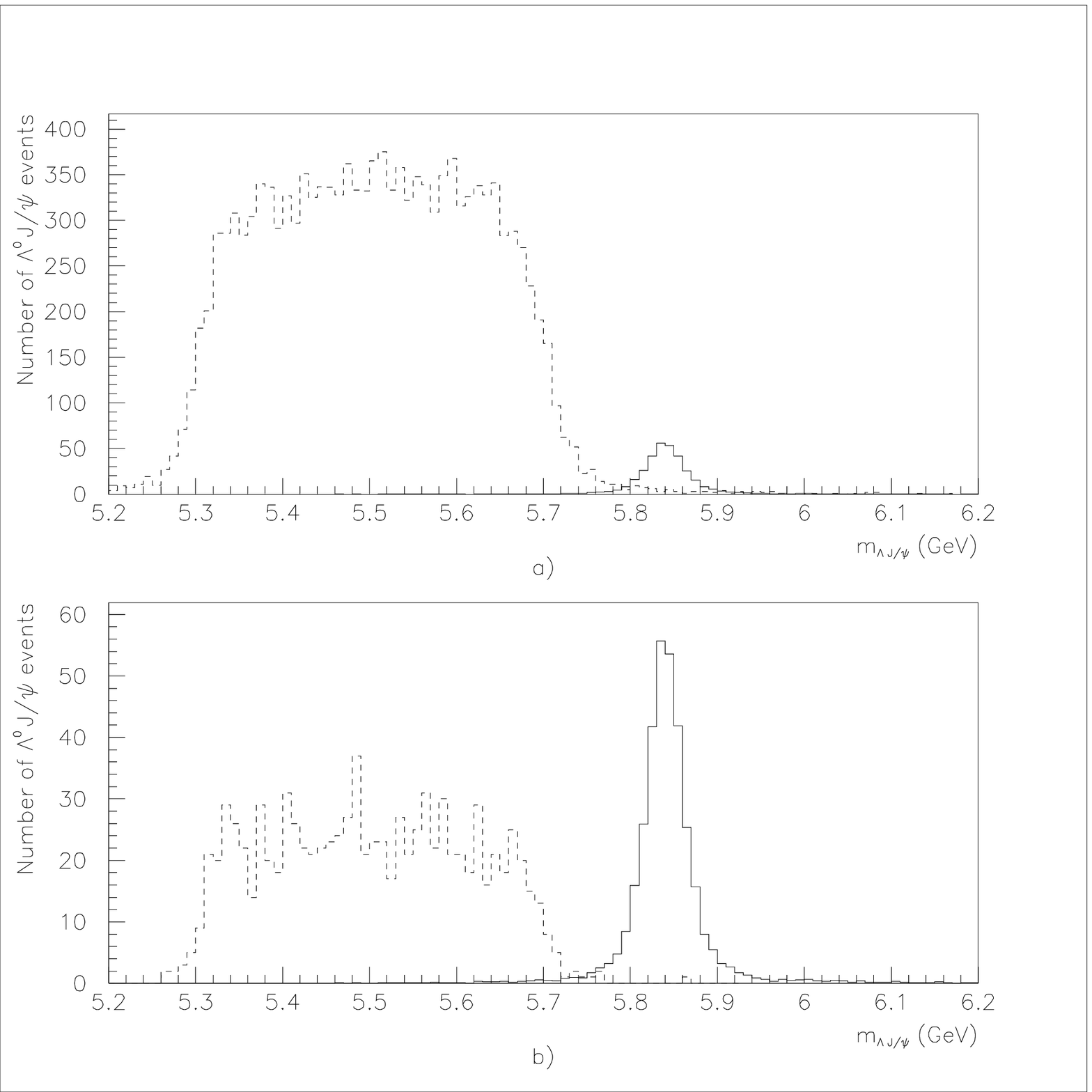}
      {\lowrm The \Lamzero $J/\psi$ effective mass distribution: The peak at
 $5.84 GeV$ is from
      \Xizerobto decay. The background with the centre at $\approx 5.5 GeV$
 comes from
    \Xinulabto , \Xizeroto and \Ximinusbto ,\Ximinusto decays (a). The events
 that passed the cut
    on the minimal distance of $J/\psi$ and \Lamzero ($d<1.5mm$) (b).}
      {obr:bckksi}
      {18cm}

%  \obr{bck_bo_mumu.eps}
%      {\lowrm Suppression of background process $B^{0}_{d}\rightarrow J\psi
%%K^{0}$
% by effective mass cut. Effective mass spectrum of $K_{0}$ when proton mass is
% assigned to one of $\pi$ (a). Effective mass of $K_{0}J/\psi$ when proton
%%mass is
% assigned to one of $\pi$ (b). The same as (b) after mass cut on \Lamzero
%%(c).}
%      {obr:bck_bo_mumu}
%      {18cm}
%

\end{document}